\documentclass{article}
\usepackage{amssymb}
\usepackage{amsmath}

\input{tcilatex}

\begin{document}

\title{Planck's constant as adiabatic invariant characterized by Hubble's
and cosmological constants}
\author{Anton Lipovka \\
Department of investigation for Physics. Sonora State University,Mexico}
\maketitle

\begin{abstract}
In the present work we suggest a non-local generalization of quantum theory
which include quantum theory as a particular case. On the basis of the idea,
that Planck constant is an adiabatic invariant of the free/coupled electromagnetic
field, we calculate the value of Plank constant from first principles,
namely from the geometry of our Universe. The basic nature of the quantum
theory is discussed. The nature of the dark energy is revealed.
\end{abstract}




\section{Introduction}

Quantum Theory (in accordance with the historical terminology, we shall call
"Quantum Theory" (QT) the theory, based on the concept of wave functions, or
probability amplitudes), that recently celebrated its 100-year anniversary,
allowed at the time to overcome a crisis that happened in atomic physics,
giving researchers a necessary tool for the calculation of atomic and
subatomic phenomena with an accuracy which is in striking agreement with
experiment. However, since its foundation and up to now, more than hundred
years ago, physicists and mathematicians are still\ trying to understand the
physical meaning of the strange QT formalism, and understand what is the
nature of quantization and most of Planck's constant?

On the one hand, Quantum Mechanics (QM) from the beginning (and then Quantum
Field Theory as its successor) was built on the axiomatic approach (which of
cource cannot be considered as an satisfactory way to construct QT). So, the
concept of the wave function was postulated for all describable entities. On
the other hand, the evolution operator for a system is linear with the wave
function, whereas its square appears as the result of the measurement
process. If we add to the above the presence of divergences and
unrenormalizability of theories in general case, the serious problems with
unification of QT and general relativity (GR), and the inability to obtain
the mass and charge from the first principles, the incompleteness of QT
becomes apparent, and thus we need to find a complete theory describing at
least the atomic and nuclear systems for the beginning.

Since the moment of discovery, QT did not please its creators, giving rise
to numerous discussions about the place for probability in physics, the
wave-particle duality, discussion of thought experiments and paradoxes. We
shall not discuss here again the well-known history of QT, for that the
reader should refer to the monograph by M. Jammer (1967). Such an "unusual"
physics, researchers had to tolerate nearly a century, excusing its numerous
defects, because QT allows calculate physically interesting phenomena in
excellent agreement with the experiment. The situation began to change in
the last decade of the 20th century, when the crisis that hit theoretical
physics became obvious to many physicists and people started talking loudly
about the problems that arise when we trying to unify QT and GR.

Among the most serious problems of the Standard Model we could mention the
following:

1. The problem of the collapse of the wave function (the problem of the
observer, or Einstein -- Podolsky -- Rosen paradox).

2. The presence of unrenormalizable (in general) divergences.

3. The huge discrepancy between the calculated with QFT methods and observed
cosmological constant (so called dark energy problem).

4. QT conflicts with general relativity at the horizon of black holes.

5. Recent experimental data obtained with the Planck satellite, which
disfavors all the best-motivated inflationary scenarios (A. Ljjas, P.J.
Steinhardt, A. Loeb 2013).

6. Inability of a reasonable harmonization or unification of the standard
model with gravity.

This incomplete list of problems indicates very serious gaps in our
understanding of Nature. For the most part, the problems appear directly or
indirectly from a misunderstanding of the basis of the quantum theory, and
the nature of its main concepts and axioms.

The present paper is urged to fill the above mentioned gaps and to specify a
way free from the difficulties listed above. We begin with a generalization
of the quantum theory because in its present form it cannot be unified with
relativistic theory.

\bigskip

\section{Quantization}

It is well known that quantum mechanics arose from the need to explain the
experimentally observed blackbody emission spectrum and atomic spectra.
Planck was the first who propose an analytical formula to describe the
spectral energy distribution which was in excellent agreement with the
experiment. However, as it was noted by Einstein (Einstein, 1906), the way
in which Planck obtained his relation, was not quite correct, though it did
lead to the correct result. The problem was that Planck included in his
formula not only the electromagnetic field, but also oscillators associated
with the matter. As a result, in the electrodynamic part, based on Maxwell's
equations, the energy of the oscillators has a continuously varying value,
while in the statistical part the same energy is considered as a discrete
value ( was quantized).

In 1905 Einstein published the work (Einstein, 1905) in which he showed that
the emission field (without any assumptions on the matter properties)
behaves, so as if consists of separate quanta (photons), characterized by
energy $hv$. Later, in 1910 Debye (Debye, 1910) showed that Planck's formula
can be deduced for the pure radiation field, absolutely without any
assumptions on the oscillator's properties of the substance. Thus Planck's
law and all its consequences, follows from the fact that the energy of
freely propagating electromagnetic field is divided by parts proportional to 
$hv$. Recently this result was confirmed experimentally by Grangier, Roger
and Aspect (1986). This fact is the only we need in order to obtain the
Planck constant from the geometry of our Universe, so reader who is
interested in the relation between the Planck constant and geometry of our
Univerce (Hubble's and Cosmological constants) can read directly the part
\textquotedblleft Adiabatic invariant\textquotedblright . But we consider
here some consequences for the Quantum theory (in Euclidean geometry), which
follows from the properties of the electromagnetic field mentioned above.

It is known that the Bohr-Sommerfeld theory (so-called old quantum theory),
based on the adiabatic hypothesis, is founded on two quantum axioms, which
when added to the axioms of classical mechanics allows us to build a quantum
theory. These two axioms are written as:

\begin{equation}
\doint p_{k}dq_{k}=n_{k}h  \tag{1.1}
\end{equation}

\begin{equation}
E_{1}-E_{2}=h\nu  \tag{1.2}
\end{equation}

The hypothesis expressed by Sommerfeld served as the basis for the writing
of these relations. It states that in each elementary process, the action of
the atom changes by an amount equal to the Planck constant. However, if we
take into account the results obtained by Einstein and Debye, we easily
receive these postulates, as a consequence of classical mechanics, i.e. we
can construct the reasonable classical theory of emission / absorption in
lines, and the classical atomic theory without recurring to the concept
(axiom) of the wave function and the problems provoked by last one. It
should be stressed here, that the so-called \textquotedblleft new quantum
theory\textquotedblright\ also is based on the axiom, and this axiom (of the
wave functions existence) cannot be explained or reduced to real physics,
whereas the Bohr-Sommerfeld axioms can be reduced to (or obtained from)
classical physics, which provide us with a fundamental view to the basic
concept and understanding the nature of the quantum theory.

To achieve the above, it should be noted, that there are only two fields,
which are carrying out interactions at big distances $(r>10^{-11}cm)$. These
are the electromagnetic and the gravitational fields. Considering that the
interaction constant for a gravitational field is negligible in comparison
with an electromagnetic one, we can surely approve the following:

Everything that we see, feel, hear, measure, and register with detectors, is
an electromagnetic field and nothing else. That is we perceive the real
world in the form of this picture, by means of electromagnetic waves
registered by us. It is important to understand, that the electromagnetic
field acts as intermediary between the observer and the real (micro) world,
hiding from us reality (the so-called idea of existence of the "hidden
variables" in QM). In our case these hidden variables lose the mystical
meaning, becoming usual classical variables - coordinates and momenta of
particles, but which can be measured only by the electromagnetic field means.

Thus as a starting point we propose the following:

1) The electromagnetic field is the only field responsible for interaction
between objects and observer in quantum mechanics.

2) The free electromagnetic field is quantized without the need of any
assumptions about the properties of oscillators. That is the Planck's
relation of $E=h\nu $ , $P=\hbar k$ is satisfied, irrespective of the
oscillators properties (see papers of Einstein (1905), Debye (1910) and
Grangier, Roger and Aspect (1986)).

The last thesis means that there exists (and therefore can be emitted) only
the photon possessing the period $2\pi $. In other words, emission /
absorption of a photon can occur only for the whole period of movement of a
charge (in system of coordinates in which proceed the emission / absorption).

Let's consider the closed system in which charge moves periodically and with
constant acceleration. In this case the Hamilton function of the electron
does not depend explicitly on time. Let's write it down as:

\begin{equation}
H=K+U=E=const  \tag{1.3}
\end{equation}

where $K$, $U$ are kinetic and potential energy and $E$ is a total energy of
system.

Then function of Lagrange is:

\begin{equation}
L=K-U=2K-E  \tag{1.4}
\end{equation}

Let's write down action for the bounded electron:

\begin{equation}
S=\underset{0}{\overset{t}{\int }}Ld\tau =2\underset{0}{\overset{t}{\int }}%
Kd\tau -Et=S_{0}-Et  \tag{1.5}
\end{equation}

\bigskip but

\begin{equation*}
\Delta S=\underset{0}{\overset{T_{1}}{\int }}L_{1}d\tau -\underset{0}{%
\overset{T_{2}}{\int }}L_{2}d\tau =0\text{ \ ,}
\end{equation*}

where $T_{1}$ and $T_{2}$ are the periods of movement of the electron in our
system on the first and second orbits respectively. Then, considering the
equation of Hamilton-Jacobi, for two different orbits 1 and 2 we have

\begin{equation*}
\Delta S=S_{2}-S_{1}=2\underset{0}{\overset{T_{2}}{\int }}K_{2}d\tau -2%
\underset{0}{\overset{T_{1}}{\int }}K_{1}d\tau -(E_{2}T_{2}-E_{1}T_{1})=0
\end{equation*}

\bigskip However (see statements 1 and 2, mentioned above)

\begin{equation}
(E_{2}T_{2}-E_{1}T_{1})=hvT_{ph}=h  \tag{1.6}
\end{equation}

is action for a emitted / absorbed photon. Thus

\begin{equation}
2\underset{0}{\overset{T_{2}}{\int }}K_{2}dt-2\underset{0}{\overset{T_{1}}{%
\int }}K_{1}dt=h  \tag{1.7}
\end{equation}

\bigskip that actually represents the first axiom of Bohr-Sommerfeld (1.1).

Let us consider for example an electron in the central field in the
nonrelativistic limit. We have: $K=\frac{1}{2}p\overset{\cdot }{q}$ and $dt=%
\frac{dq}{\overset{\cdot }{q}}$ , where $p=-\frac{\partial H}{\partial 
\overset{\cdot }{q}}$\ . Then expression (1.7) gives

\begin{equation}
\doint p_{2}dq_{2}-\doint p_{1}dq_{1}=h  \tag{1.8}
\end{equation}

which for s-state of atom of hydrogen gives a known relation

\begin{equation}
mr_{2}^{2}\overset{\cdot }{\varphi }_{2}-mr_{1}^{2}\overset{\cdot }{\varphi }%
_{1}=\hbar \text{ ,}
\end{equation}%
\newline

or, the same

\begin{equation}
M_{2}-M_{1}=\hbar \text{ ,}  \tag{1.9}
\end{equation}

where $M_{2}$ and $M_{1}$ are the angular momenta. To write down the
expression (1.9) we used the fact that the obtained values $mr^{2}\overset{%
\cdot }{\varphi }$ formally coincides with the angular momenta for electron
in the central field.

Let's put $M_{0}=0$ (that corresponds to $r_{0}=0$). In this case we have $%
M_{1}=$ $M_{0}+\Delta M$, but $\Delta M=\hbar $, so we obtain

\begin{equation}
M_{1}=M_{0}+\hbar =\hbar \text{ , }M_{2}=M_{1}+\hbar =2\hbar \text{..., }%
M_{n}=n\hbar  \tag{1.10}
\end{equation}

From expression (1.10) and a principle of mechanical similarity for the
central potentials of $U\sim r^{k}$ , we have

\begin{equation}
\frac{M^{\prime }}{M}=\left( \frac{r^{\prime }}{r}\right) ^{1+\frac{k}{2}};%
\frac{E^{\prime }}{E}=\left( \frac{r^{\prime }}{r}\right) ^{k}\text{ ,}
\end{equation}

from where follows:

\begin{equation}
r_{n}=r_{1}\left( n\right) ^{\frac{1}{1+\frac{k}{2}}}\text{ and \ }%
E_{n}=E_{1}\left( n\right) ^{\frac{k}{1+\frac{k}{2}}}  \tag{1.11}
\end{equation}

Then for a classical harmonic oscillator ($k=2$.) from (1.11) we get:

\begin{equation}
r_{n}=r_{1}\sqrt{n}\text{ ; }E_{n}=E_{1}n  \tag{1.12}
\end{equation}

\bigskip and for atom of hydrogen

\begin{equation}
r_{n}=r_{1}n^{2}\text{ ; }E_{n}=\frac{E_{1}}{n^{2}}  \tag{1.13}
\end{equation}

The value $E_{1}$ in the last expression can be found easily from expression
(1.6) $\left( E_{2}T_{2}-E_{1}T_{1}\right) =h$ . \bigskip Accepting
classical value of the period

\begin{equation}
T=\pi e^{2}\sqrt{\frac{m}{2\left\vert E\right\vert ^{3}}}\text{ ,} 
\tag{1.14}
\end{equation}

and taking into account (1.13) $E_{2}=\frac{1}{4}E_{1}$ we have: \ 

\begin{equation}
E_{1}=\frac{me^{4}}{2\hbar ^{2}}  \tag{1.15}
\end{equation}

Thus we showed that so-called quantization of system (axioms of
Bohr-Sommerfeld) arises in absolutely classical way from the intrinsic
properties of the electromagnetic field and cannot be treated as quantum
property of space or matter.

\section{Harmonic oscillator}

\bigskip

There is a common misconception that the additional term of $1/2$, which
appears in the energy of the harmonic oscillator, is a quantum effect and is
associated with the so-called zero - oscillations. Due to the methodological
importance of this question, we discuss it here in a little more detail in
the non-relativistic limit, and show that it is a purely classical effect.

Accordingly to classical mechanics, the energy of the harmonic oscillator is:

\begin{equation}
E=\frac{m}{2}(\ \overset{\cdot }{r}^{2}+\omega ^{2}r^{2})  \tag{2.1}
\end{equation}%
where $\omega =\sqrt{\frac{k}{m}}$. Then, considering that for the harmonic
oscillator $\overset{\_}{T}=\overset{\_}{U}$ , we obtain for the average
energy for the period:

\begin{equation}
E_{n}=mr_{n}^{2}\omega ^{2}  \tag{2.2}
\end{equation}

To carry out transition from an initial state of system to the final one $%
E_{n\text{ }}\rightarrow E_{k}$, we should "take away" energy from our
oscillator by electromagnetic field.

It is known that emission of an electromagnetic field by a moving charge
differ from zero only if we integrate for the complete period $T$ of
movement in the course of which the emission or absorption appears. It
corresponds to the fact that the complete photon instead of its part is
emitted / absorbed, that is the generated field satisfying to a periodicity
condition.

The factor of proportionality between energy and frequency for a free
electromagnetic field is $\hbar $:

\begin{equation}
\Delta E=E_{n\text{ }}-E_{k}=\hbar \omega _{nk}  \tag{2.3}
\end{equation}%
(Once again we emphasize here that as it follows from Einstein's and Debye
works, the constant $\hbar $ concerns only to the electromagnetic field and
do not appears in any way from matter properties, or the size of the system
under consideration). Expression (1.12) gives a ratio between energy levels,
however considering (2.3) it is clear that the residual energy $%
E_{_{0}}=U(r_{1})$ cannot be emitted by a photon $\hbar \omega $, because

\begin{equation}
\Delta E=E_{1\text{ }}-E_{0}=mr_{1}^{2}\omega ^{2}-\frac{1}{2}%
mr_{1}^{2}\omega ^{2}=\frac{1}{2}E_{1\text{ }}<\hbar \omega _{{}}  \tag{2.4}
\end{equation}

Therefore this additive constant (which appears owing to the shape of the
potential) should be simply added to the expression (1.12):

\begin{equation}
E_{n}=nE_{1}+\frac{1}{2}E_{1\text{ }}=\hbar \omega (n+\frac{1}{2})  \tag{2.5}
\end{equation}

Thus, the additive constant $1/2$ appears naturally from classical
consideration.

\section{Quantum mechanics is the Fourier - transformed classical mechanics}

\bigskip

In standard textbooks of quantum mechanics problems arise and are solved for
isolated systems, while the periodic electromagnetic field is not included
in the Hamiltonian of the system under consideration. For example a harmonic
oscillator, the hydrogen atom, molecular potentials, etc. Thus on the one
hand any changes in the system (transitions between levels) are associated
with the photons, but on the other hand, the free / bounded electromagnetic
field in such Hamiltonian does not appears. Reasonable questions arise:
where the electromagnetic field is and why it does not appears in the
Hamiltonian $H$? How these electromagnetic fields are taken into account for
the quantization of such systems?

At the beginning of the 20-th century, the equations describing the quantum
system have been intuitively guessed and accepted for the calculations
(despite the emerging issues), because calculated results were consistent
with the experiments at the time. However, the meaning of the wave equations
and the wave function itself is still not completely understood. In this
section we will show sense of the formalism of quantum mechanics making a
start from bases of classical mechanics

For simplicity, consider the one-dimensional motion. The generalization to
three dimensions is obvious. Suppose we have the classical equation for
energy:

\begin{equation}
H=E  \tag{3.1}
\end{equation}

Here $H$ - classical Hamilton function of the system and $E$ - total energy
of the system. Let's consider a particle in the field of $U(x)$. For a total
energy of system we have two possibilities:

1) $E<0$, the system is bounded, we have a periodic movement,

2) $E>0$, the system is unbounded, we have a free movement.

Virtual photon (strictly speaking we should say \textquotedblleft coupled
with electron electromagnetic field, part of which can be
emitted\textquotedblright ), can be described by harmonic function:

\begin{equation}
\varphi =\exp (-ik_{\alpha }x^{\alpha })  \tag{3.2}
\end{equation}

where $k_{\alpha }$ and $x_{_{^{\alpha }}}$ are 4 -vectors.

Consider $E<0$, that corresponds to a discrete spectrum in quantum
mechanics. The case of continuous spectrum, when $E>0$, differs only by
replacement of sums by integrals, but the entire derivation of the equations
is done similarly.

Let's apply to (3.1) the opposite Fourier - transform on coordinate x by
using harmonic function of the virtual photon:

\begin{equation}
\int H(k,x)\varphi (k,x)dx=\ \ \int E\varphi (k,x)dx  \tag{3.3}
\end{equation}%
\ \ \ \ \ \ \ \ \ \ \ \ \ \ \ \ \ \ \ \ \ \ \ \ \ \ \ \ \ \ \ \ \ \ \ \ \ \
\ 

or%
\begin{equation}
\int \frac{p^{\mathbf{2}}}{2m}\ e^{-\frac{i}{\hbar }(px-Et)}dx+\int U(x)e^{-%
\frac{i}{\hbar }(px-Et)}dx=\int E\ e^{-\frac{i}{\hbar }(px-Et)}dx  \tag{3.4}
\end{equation}%
\ \ \ 

from where obtain:%
\begin{equation}
\int dx\left[ -\frac{\hbar ^{\mathbf{2}}}{2m}\frac{\partial ^{\mathbf{2}}}{%
\partial x^{\mathbf{2}}}+U(x)=-i\hbar \frac{\partial }{\partial t}\right]
e^{-\frac{i}{\hbar }(px-Et)}  \tag{3.5}
\end{equation}%
\ \ \ \ \ \ \ \ \ \ \ \ \ \ \ \ \ \ \ \ \ \ \ \ \ \ \ 

or

\begin{equation}
\int dx\left[ (\hat{H}-E)\varphi =0\right]  \tag{3.6}
\end{equation}

where $\hat{H}$ is the Hamilton operator of the system under consideration.

We note here that the replacement of an electron by a positron (formally
changes the sign in the exponent for opposite one), leads to the replacement
of $t$ by $-t$ in equation (3.5). In equation (3.6) in the brackets we have
the Hamilton operator $\hat{H}$, which actually is the Liouville operator,
so it has a complete set of orthogonal eigenfunctions.

Let $\Psi _{k}$ (x) be a complete set of eigenfunctions of the operator $%
\hat{H}$, then we can write down

\begin{equation}
\varphi (p,x)=\underset{m}{\dsum }a_{m}(p)\Psi _{m}(x)  \tag{3.7}
\end{equation}%
\ \ \ \ \ \ \ \ \ \ \ \ \ \ \ \ \ \ \ \ \ \ \ \ \ \ \ \ \ \ \ \ \ \ \ \ \ \
\ \ \ \ \ \ \ \ \ \ \ \ \ \ \ \ \ \ \ \ \ \ \ 

and the equation (3.6) becomes%
\begin{equation}
\dint dx\underset{m}{\dsum }a_{m}(p)\left[ (\hat{H}-E)\Psi _{m}=0\right]  
\tag{3.8}
\end{equation}%
\ \ \ \ \ \ \ \ \ \ \ \ \ \ \ \ \ \ \ \ \ \ \ \ \ \ \ \ \ \ \ \ \ \ \ \ \ \
\ \ \ \ \ \ \ \ \ 

or (taking into account that $\Psi _{m}$ are eigenfunctions of the Liouville
operator $\hat{H}$), the expression in square brackets is:

\begin{equation}
\hat{H}\Psi _{m}(x)=E_{m}\Psi _{m}(x).  \tag{3.9}
\end{equation}

This is the equation of Schr\"{o}dinger in coordinate representation. It is
clear that if in (3.3) we integrate over $p$ instead of coordinate, in the
same way we will obtain the Schr\"{o}dinger equation, but now in $p$ -
representation.

\begin{equation}
\hat{H}\Psi _{m}(p)=E_{m}\Psi _{m}(p)  \tag{3.10}
\end{equation}%
\ \ \ \ \ \ \ \ \ \ \ \ \ \ \ \ \ \ \ \ \ \ \ \ \ \ \ \ \ \ \ \ \ \ \ \ \ \
\ \ \ \ \ \ \ \ \ \ \ \ \ \ \ \ \ \ \ \ \ \ \ \ \ \ \ \ \ \ \ \ \ \ \ \ 

Let's make now inverse transformation of expression (3.8). We have:%
\begin{equation}
\diint dx\underset{m}{\dsum }\varphi ^{\ast }(k,x)a_{m}\left[ \hat{H}\Psi
_{m}-E\Psi _{m}\right] dp=0  \tag{3.11}
\end{equation}%
\ \ \ \ \ \ \ \ \ \ \ \ \ \ \ \ \ \ \ \ \ \ \ \ \ \ \ \ \ \ \ \ \ \ \ \ \ \
\ \ \ \ \ \ \ \ 

considering that

\begin{equation}
\varphi ^{\ast }(k,x)=\underset{n}{\dsum }a_{n}^{\ast }(p)\Psi _{n}^{\ast
}(x)\ \ \ \   \tag{3.12}
\end{equation}

we can obtain

\begin{equation}
\diint dxdp\underset{m}{\dsum }\underset{n}{\dsum }a_{m}a_{n}^{\ast }\Psi
_{n}^{\ast }(x)\left[ \hat{H}-E\right] \Psi _{m}(x)=0  \tag{3.13}
\end{equation}

or in another form:%
\begin{equation}
\dint dp\underset{m}{\dsum }\underset{n}{\dsum }a_{m}a_{n}^{\ast }\langle
\Psi _{n}^{\ast }\mid \left[ \hat{H}-E\right] \mid \Psi _{m}\rangle =0 
\tag{3.14}
\end{equation}

Which immediately implies matrix notation of quantum mechanics.

So, we have showed that :

\qquad 1) The quantum mechanics is the Fourier - transformed classical
mechanics, and transformation goes on the function of the electromagnetic
field coupled with electron, which cannot enter obviously into the Schr\"{o}%
dinger equations, remaining out of consideration framework.

\qquad 2) The quantum theory is an incomplete (local) theory because it is
based on an incomplete (local) Schr\"{o}dinger equation (3.9) instead of the
complete (non-local) equation (3.8) where the virtual electromagnetic field
appears as coefficients $a_{m}(p)$ under summation and integration.

Thus so-called wave functions are not "probability density" but are just
eigenfunctions of the operator of Liouville which forms the problem of
Sturm-Liouville. And this is the eigenfunctions that allow us to make
decomposition of the electromagnetic field coupled with a charge. It should
be stressed here, the theory based on the equation (3.8) do not suffer of
the wave function collapse problem, and the Einstein -- Podolsky -- Rosen
paradox does not appears. In consequence with the expression (3.8), the wave
function is defined completely by the photon, and its collapse appears
within the volume $\lambda ^{3}$ , where $\lambda $\ is the wavelength of
the photon (see integration on $dx$ in (3.8)). So, within this complete
theory there do not appear movements characterized by velocities more than
light speed, as it take place in the Einstein -- Podolsky -- Rosen paradox
for the Schr\"{o}dinger equation (3.9).

To conclude, the uncertainty principle $\Delta p\Delta x$ $\sim $ $\hbar $
should be mentioned briefly. As it was mentioned above, any measurement
occurs with the assistance of a photon. In this way, we can measure the
coordinates of the object with the precision of up to $\Delta x=\lambda
/\cos \varphi $\ where $\lambda $\ is wavelength of the photon. However in
the course of coordinate measurement the photon transfers a part of their
impulse to the measured object so we can write $\Delta p=\hbar k\cos \varphi 
$\ . \ Combining the first expression with the second one we have $\Delta
p\Delta x$ $\sim $ $\hbar $\ .

On the other hand, the phase is an invariant, so we can conclude that
symmetric expression also take place \ $\Delta E\Delta t$ $\sim $ $\hbar $.

\section{Adiabatic invariant}

\bigskip

From astronomical observations it is well established that we live in a
non-stationary Universe, in which all parameters change over time. By taking
into account this fact, let's consider an isolated mechanical system makes
finite movement. Without loss of a generality we consider only one
coordinate $q$, characterizing movement of the system. Suppose also that
movement of the system is characterized by a certain parameter $r$ . Here we
can take $r=r_{u}$- radius of the Universe or $r=R$ - scalar curvature of
space. The final result will not depend on our choice. Let the parameter $r$
adiabatically change with time, i.e.

\begin{equation}
T\ll \frac{r}{\overset{\cdot }{r}}  \tag{4.1}
\end{equation}%
\bigskip where $T$ - is the characteristic time, or period of motion of our
system. From this relation one can obtain an estimation for the proper
frequency of the system satisfying the adiabatic condition:

\begin{equation*}
\nu \gg 10^{-18}\text{ }\left[ s^{-1}\right] ,
\end{equation*}%
which actually corresponds to the always fulfilled relation $\lambda
_{ph}\ll r_{u}$ (the wavelength of a photon is much less than size of the
Universe). It is clear that the system in question (photon) in this case is
not isolated, and for the total system energy we have the linear
relationship $\overset{\cdot }{E}\sim \overset{\cdot }{r}$. The Hamiltonian
of the system in this case depends on parameter $r$, therefore

\begin{equation}
\overset{\cdot }{E}=\frac{\partial H}{\partial t}=\frac{\partial H}{\partial
r}\frac{\partial r}{\partial t}  \tag{4.2}
\end{equation}%
Averaging this expression on the period, we obtain

\begin{equation}
\doint \left( \frac{\partial p}{\partial E}\overline{\frac{\partial E}{%
\partial t}}+\frac{\partial p}{\partial r}\frac{\partial r}{\partial t}%
\right) dq=0  \tag{4.3}
\end{equation}%
or designating our adiabatic invariant as $h$ , get from (4.3)

\begin{equation}
\overline{\frac{\partial h}{\partial t}}=0  \tag{4.4}
\end{equation}%
where 
\begin{equation}
h=\frac{1}{2\pi }\doint pdq  \tag{4.5}
\end{equation}%
is the Planck's constant on their sense. Considering that%
\begin{equation}
2\pi \frac{\partial h}{\partial E}=\doint \frac{\partial p}{\partial E}dq=T 
\tag{4.6}
\end{equation}%
we can write down the energy of a photon as

\begin{equation}
E=h\nu +E_{0}  \tag{4.7}
\end{equation}

It should be noted here, that $E_{0}\neq 0$ for general case.

\section{Relation between the geometry of the Universe and the value of
Planck constant.}

\bigskip

Earlier we have shown how the quantum mechanical picture of surrounding
reality appears. In the present section we obtain the important quantitative
characteristic of the quantum theory - value of the Planck constant, from
observable geometry of the Universe.

It is well known that General Relativity formulated on Riemann manifold has
some difficulties. Among the most significant the following should be
mentioned:

1. The presence of singularities.

2. Inability to take into account the "large numbers" of Eddington-Dirac
which formally suggest a relation between cosmological and the quantum
values.

3. The cosmological constant which has no explanation within the framework
of GR and QFT.

To search for a solution of these problems we must consider more general
extensions of the Riemann geometry. One of its possible natural extensions
is the geometry of Riemann - Cartan in which the theory of Einstein - Cartan
with asymmetrical connections can be developed. There is a variety of
reasons for such a choice:

1. The theory of Einstein - Cartan satisfies the principle of relativity and
also the equivalence principle and does not contradict the observational
data.

2. It follows necessarily from gauge theory of gravitation.

3. It is free from the problem of singularities.

4. It suggests the most natural way to explain the cosmological constant as
a non-Riemannian part of the scalar curvature of space, caused by torsion.

Within Riemann's geometry, as it is known, for the tensor of electromagnetic
field we have:

\begin{equation}
A_{\nu ;\mu }-A_{\mu ;\nu }=A_{\nu ,\mu }-A_{\mu ,\nu }  \tag{5.1}
\end{equation}

(due to the symmetry of connections, the covariant derivatives of
4-potencial in the field tensor can be substituted by partial derivatives).
But in the case of Einstein -- Cartan theory with asymmetrical connections,
the ratio (5.1) is not more fulfilled and an additional term in the tensor
of electromagnetic field appears.

To construct a theory we need the Lagrangian, which includes a natural
linear invariant - the scalar curvature obtained by reduction of the Riemann
- Cartan tensor of curvature. Let's accept from the beginning that curvature
of space is small (that conforms to experiment) and, therefore, in approach
interesting for us we can neglect by quadratic invariants in Lagrangian,
having written down action for a gravitational field and a matter in
Riemann-Cartan geometry this manner:

\begin{equation}
S=S_{g}+S_{m}=\frac{c^{3}}{16\pi G}\underset{\Omega }{\int }\overset{%
\backsim }{R}\sqrt{-g}d\Omega +\frac{1}{c}\underset{\Omega }{\int }\overset{%
\backsim }{\mathcal{L}}_{m}\sqrt{-g}d\Omega  \tag{5.2}
\end{equation}

Here $c$ -- light velocity, $G$ -- gravitational constant, $g$ --
determinant of the metric tensor $g^{\alpha \beta }$, $\overset{\backsim }{R}
$ is scalar curvature and $\overset{\backsim }{\mathcal{L}}_{m}$ is the
Lagrangian of the matter which have been written down for Riemann-Cartan
manifold, $d\Omega =d^{4}x$. Varying it we obtain

\begin{equation}
\delta S_{g}=-\frac{c^{3}}{16\pi G}\underset{\Omega }{\int }\left( \overset{%
\backsim }{R}_{\alpha \beta }-\frac{1}{2}g_{\alpha \beta }\overset{\backsim }%
{R}\right) \delta g^{\alpha \beta }\sqrt{-g}d\Omega  \tag{5.3}
\end{equation}

\bigskip and%
\begin{equation}
\delta S_{m}=\frac{1}{2c}\underset{\Omega }{\int }\overset{\backsim }{T}%
_{\alpha \beta }\delta g^{\alpha \beta }\sqrt{-g}d\Omega  \tag{5.4}
\end{equation}

\bigskip or

\begin{equation*}
-\frac{c^{3}}{16\pi G}\underset{\Omega }{\int }\left( \overset{\backsim }{R}%
_{\alpha \beta }-\frac{1}{2}g_{\alpha \beta }\overset{\backsim }{R}-\frac{%
8\pi G}{c^{4}}\overset{\backsim }{T}_{\alpha \beta }\right) \delta g^{\alpha
\beta }\sqrt{-g}d\Omega =0
\end{equation*}

and finally 
\begin{equation}
\overset{\backsim }{R}_{\alpha \beta }-\frac{1}{2}g_{\alpha \beta }\overset{%
\backsim }{R}=\frac{8\pi G}{c^{4}}\overset{\backsim }{T}_{\alpha \beta } 
\tag{5.5}
\end{equation}

Here $\overset{\backsim }{T}_{\alpha \beta }$is a tensor of density of
energy - momentum of the matter in space with geometry of R-C. Simplifying
on indexes we have:

\begin{equation*}
\overset{\backsim }{R}=-\frac{8\pi G}{c^{4}}\overset{\backsim }{T}
\end{equation*}%
\ 

\bigskip or in other form

\begin{equation}
\left( R-4\Lambda \right) =-\frac{8\pi G}{c^{4}}\overset{\backsim }{T}\text{
\ \ ,}  \tag{5.6}
\end{equation}

where ~$R$\ - is the scalar formed of the Riemann's tensor, $\Lambda =(R-%
\overset{\backsim }{R})/4$\ and $\overset{\backsim }{T}$\ - trace of tensor $%
\overset{\backsim }{T}_{\alpha \beta }$ of electromagnetic field in R-C
geometry.

In the right side of (5.6) we have the value associated with the difference
of geometry from the Riemann one (the trace of a tensor $T_{\alpha \beta }$
for the electromagnetic field is equal to zero in Riemann's geometry because
of symmetry of connections) that we want to evaluate. The problem in the
direct estimation of the value of $\overset{\backsim }{T}$\ is that we do
not know the true metric of the Universe we live in. We also do not know the
real connection coefficients of our space. For this reason, we cannot
directly calculate the value that we are interested in. Accordingly, we
cannot just write out a corresponding amendment to the energy of
electromagnetic field. However we can estimate this value indirectly,
considering that the left part of expression (5.6) contains observable
values.

As follows from the section "adiabatic invariant", for the action of
electromagnetic field we have:

\begin{equation}
S=S_{0}-h  \tag{5.7}
\end{equation}%
where $S_{0}$ is an integration constant and $h$ - is the adiabatic
invariant caused by slowly changing curvature of space in the Riemann-Cartan
Universe. Then, considering that the trace of the tensor $T_{\alpha \beta }$
for the electromagnetic field is equal to zero in Riemann's geometry, we can
write at once from (5.6)

\begin{equation}
\left( R-4\Lambda \right) \frac{c^{4}}{8\pi G}\approx 2\frac{h}{\Delta t_{0}}%
=2h\nu _{0}  \tag{5.8}
\end{equation}%
We emphasize here that on the left side of this expression, we have the
observed quantities which characterize the Universe geometry, while on the
right side, appears the Planck constant, which in turn, characterize a
microcosm. The value $\Delta t_{0}$ is minimum possible interval of time
corresponding to action $h$. To find it we notice that energy of
corresponding electromagnetic field can change only by the value $h\nu $.
(see first part of the paper). Let's consider as an example the atom of
hydrogen (for our purposes we could consider any system). The first Bohr
orbit is characterized by value $M_{1}=m_{e}a_{0}V_{0}=\hbar $ , where $m_{e}
$ is the electron mass, $a_{0}$ -- Bohr radius and $V_{0}$ - velocity of the
electron at first Bohr orbit. State with $M_{0}=0$ is not achievable for our
system. As radius reduces from $a_{0}$ to the Compton wavelength $\lambda
_{c}/2\pi $ , the value $M_{1}=\hbar $ cannot be changed, for the photon
cannot be emitted. So we can write $\lambda _{c}c/2\pi =a_{0}V_{0}$, or $\nu
_{0}=1/\Delta t_{0}=c/a_{0}=2\pi V_{0}/\lambda _{c}=5.6652\times
10^{18}[s^{-1}]$\ . Here we need to emphasize especially that time, as well
as space, are continuous, i.e. they do not quantized. The interval $\Delta
t_{0}=1.7651\times 10^{-19}[s]$ is the minimum interval of time,
corresponding to value $h$. From expression (5.8), we can write

\begin{equation}
\left( R-4\Lambda \right) \frac{c^{3}a_{0}}{16\pi G}\approx h  \tag{5.9}
\end{equation}%
where

\begin{equation}
R=4\pi ^{2}\frac{H_{0}^{2}}{c^{2}}.  \tag{5.10}
\end{equation}

Let's estimate the Planck constant. The measured values of the Hubble
constant were presented in works (Riess et al. 2009) $H_{0}=74.2\pm 3.6$ $%
[km $ $s^{-1}Mpc^{-1}]$\ and (Riess et al. 2011) $H_{0}=73.8\pm 2.4$\ $[km$ $%
s^{-1}Mpc^{-1}]$. Let's take for our assessment average value $H_{0}=74$ $%
[km $ $s^{-1}Mpc^{-1}]$. Cosmological constant $\Lambda $ we adopt according
to measurements $\Omega _{\Lambda }=0.7$ and we accept critical density $%
\rho _{c}=1.88\times 10^{-29}[g$ $cm^{-3}]$\ . Then, from expression (5.9)
we obtain value for the Planck's constant $h=6.6\times 10^{-27}[erg$ $s]$,
that coincides to within the second sign with experimental value.

Recently, the issue of a possible change of the fine structure constant $%
\alpha $ on time is widely debated, so for convenience, we put here another
interesting relationship, which follows from (5.9)

\begin{equation}
\left( R-4\Lambda \right) \frac{c^{4}}{16\pi G}=2\pi m_{e}c^{2}\alpha 
\tag{5.11}
\end{equation}

\bigskip

\section{Other observational effects}

\bigskip

The results suggested in present work can be proved by independent
experiments. The most basic of them is of course the double slit experiment.
Recently it was accurately carried out by Demjanov (2010), which clearly
argued for our model of non-local quantum theory. Another possible
experiment could be a measurement of the blackbody spectrum in far
Reyleigh-Jeans region. As it was shown earlier, if the geometry of
Riemann-Cartan has non-zero scalar curvature, in expression for energy of
electromagnetic field appears $h\nu $ . The energy of one photon in this
case is:

\begin{equation}
E_{\nu }=E_{\nu }^{0}+h\nu  \tag{6.1}
\end{equation}%
where $\nu $ is a frequency of a photon, and $E_{\nu }^{0}$ is a small
aditional energy:

\begin{equation}
E_{\nu }^{0}=\frac{1}{16\pi }\int (E^{2}+H^{2})dV-h\nu \ \ .  \tag{6.1}
\end{equation}

Integration here is carried out over the volume of one photon. Intensity of
the black body emission in this case one can write as

\begin{equation}
B_{\nu }=(E_{\nu }^{0}+h\nu )\frac{2\nu ^{2}}{c^{2}}\frac{1}{\exp \{\frac{%
E_{\nu }^{0}+h\nu }{kT}\}-1}  \tag{6.3}
\end{equation}

As one can see, in Wien and in the near Reyleigh-Jeans region the spectrum
is almost coincide with Planck one because the value $E_{\nu }^{0}$ \ is
small. However it is clear that the small additive energy $E_{\nu }^{0}$ can
lead to some deviations from Planck spectrum in far Reyleigh-Jeans region
and, probably, such deviation could be measured experimentally. It is
necessary to emphasize that such experiment has independent great importance
because will allow to state an assessment to the value $E_{\nu }^{0}$ and to
throw light on the geometrical nature of electromagnetic field.

\bigskip

\section{\protect\bigskip Conclusion}

In present work we made the next logical step towards implementation of the
program begun by Einstein and Schr\U{446}dinger in the fifties of the XX
century (model of Einstein -- Cartan - Schr\"{o}dinger). Namely we show that
the Planck constant is actually the adiabatic invariant of the
electromagnetic field, characterized by scalar curvature of space of the
Riemann -- Cartan geometry. The main results of present work are:

\qquad 1) For the first time we obtained the ratio between Riemannian scalar
curvature of the Universe, the Cosmological constant and Planck's constant
(see expression (5.9)), true up to the second sign ($\Omega _{\Lambda }\gg
\Omega _{b}$).

\qquad 2) It is stressed that due to change of geometry of the Universe, the
Planck constant changes on time too.

\qquad 3) The physical sense of the cosmological constant, as the no -
Riemannian part of the scalar curvature, which appears due to the presence
of torsion (asymmetrical connections), is revealed for the first time.

\qquad 4) Dependence of the fine structure constant on the total scalar
curvature of the Universe is obtained (5.11).

\qquad 5) Within the used framework, natural unification of gravitation with
the quantum theory is obtained.

\qquad 6) In linear on curvature invariants approach, the spectral density
of the blackbody radiation is obtained.

\qquad 7) Bases of the quantum theory are reconsidered and the physical
sense of wave function is found. It is shown that if we eliminate an
unnatural axiom of existence of wave function, the huge discrepancy between
calculated by the QFT methods and observed cosmological constant, disappears.

\qquad 8) The approach based on the equation (3.8), completely removes a
problem of collapse of wave function and classically resolves the Einstein
-- Podolsky -- Rosen paradox. According to (3.8) "wave function", as it
should be, is completely determined by correspponding electromagnetic field
and its collapse occurs at scales of wavelength of the photon (integration
on $dx$ in (3.8)). Thus in this way there is no need for transmission of a
signal with a superlight speed as it takes place in paradox of E-P-R for the
Schr\"{o}dinger equation (3.9).

\section{Acknowledgements}

I would like to acknowledge Dr. J. Saucedo for the valuable criticism and
comments. I also grateful to the Pulkovo Observatory team and particularly
to Dr. E. Poliakow for the opportunity to spent part of 2008 year in the
Observatory.

\bigskip

\section{\protect\bigskip Bibliography}

Debye, P. 1910. Der Wahrscheinfichkeitsbegriff in der Theorie der Strahlung.
Annalen der Physik. 33, 1427 - 1434.

Demjanov, V.V. 2010. Experiments performed in order to reveal fundamental
differences between the diffraction and interference of waves and electrons.
arXiv: 1002.3880

Einstein, A. 1905. Uber einen die Erzeugung und Verwandlung des Lichtes
betreffenden heuristischen Gesichtspunkt. Annalen der Physik. 17, 132-148. ;
Einstein, A. 1965. On a heuristic view point concerning the production and
transformation of light. American Journal of Physics. 33, 367-374.

Einstein, A. 1906. Zur Theorie der Lichterzengung und Lichtabsorption.
Annales der Physik, 20, 199 -- 206.

Grangier, P., Roger, G., Aspect, A.. Europhys. Lett. Vol.1. Pp. 173-179,
1986.

Jammer, M. 1967. The conceptual development of quantum mechanics. Mc Graw
Hill. New York.

Ljjas, A., Steinhardt, P.J., Loeb, A. 2013. Inflationary paradigm in trouble
after Planck2013. arXiv: 1304.2785.

Riess A.G., Macri L., Casertano S., et al. 2009. A Redetermination of the
Hubble Constant with the Hubble Space Telescope from a Differential Distance
Ladder. Ap.J. 699, 539-563

Riess A.G., Macri L., Casertano S., et al. 2011. A 3\% Solution:
Determination of the Hubble Constant with the Hubble Space Telescope and
Wide Field Camera 3. Ap.J. 730, 119-136.

\end{document}